\documentstyle[amssymb]{article}

\newtheorem{rem}{Remark}
\newtheorem{rems}[rem]{Remarks}

\newcommand{\Rn}{{\rm I\!R}} 
\newcommand{\Cn}{{\setbox0=\hbox{
$\displaystyle\rm C$}\hbox{\hbox
to0pt{\kern0.6\wd0\vrule height0.9\ht0\hss}\box0}}} 
\newcommand{\idty}{\hat{\rm 1\mskip-4mu l}} 
\newcommand{\Zn}{{\hbox{$\sf\textstyle Z\kern-0.3em Z$}}} 

\newcommand{\cA}{{\cal A}}
\newcommand{\cB}{{\cal B}}

\newcommand{\cE}{{\cal E}}

\newcommand{\cF}{{\cal F}}
\newcommand{\cH}{{\cal H}}
\newcommand{\cI}{{\cal I}}
\newcommand{\cL}{{\cal L}}
\newcommand{\cM}{{\cal M}}

\newcommand{\cP}{{\cal P}}

\newcommand{\cO}{{\cal O}}

\newcommand{\cS}{{\cal S}}

\begin{document}
\title{On Kolgomorov-Sinai entropy and its quantization
    \thanks{ 
\ \ Work supported by KBN grant PB/0273/PO3/99/16 
}}
\author{
W\ A\ Majewski 
    \thanks{\ \ Institute of Theoretical Physics and Astrophysics, University
of Gda\'nsk, Wita Stwosza 57, PL 80-952 Gda\'nsk, Poland.  E-mail: fizwam@univ.gda.pl}
}
\date{}
\maketitle{}


\begin{abstract}
In this paper we present the new approach to Kolgomorov-Sinai entropy and its quantization.
Our
presentation stems from an application of the Choquet theory to the theory of 
decompositions of states and therefore, it resembles our  
rigorous description of entanglement of formation. 
\end{abstract}

\section{Introduction}
The problem of quantization of dynamical entropy 
(so Kolmogorov-Sinai entropy \cite{Kol}, \cite{S}) has attracted
much attention and that concept has been widely considered in different 
mathematical and physical contexts (cf. \cite{ConSt}, \cite{CNT} and references therein, 
see also \cite{Savagaut}, \cite{AF}, \cite{Voicu}, \cite{GoldPen}). 
Though the concept of dynamical entropy has so many quantum counterparts
it seems that, frequently, they have undesired properties. 

In this paper we are concerned with the algebraic reformulation of original
Kolmogorov-Sinai (KS) entropy in such a way that its quantization is natural
and straightforward. To this end, firstly  
 we look more closely at the original definition KS entropy from 
the pure $C^*$-algebraic point of view. Namely,
there is a difficulty in implementing the definition 
of partition for the non-commutative case.
 To overcome this problem in another way to that of given in \cite{CNT} or in \cite{AF},
and to get a well defined function of dynamical system with nice properties we shall
use the theory of decomposition which is based on the theory
of compact convex sets and boundary integrals. Let us note that this strategy
proved to be very fruitful in the recent analysis of quantum entanglement and
quantum correlations (cf. \cite{M}, \cite{M1}).  
Then, having a reformulation of
Kolmogorov-Sinai entropy we will discuss the question of its quantization.
The paper is organized as follows. In Section II we set up notation
and terminology, and we review some of the standard facts on the theory
of decomposition. Section III contains our description of KS entropy.
In section IV we present our version of quantization of dynamical entropy
while the final section V contains some concluding remarks.

\section{Preliminaries}

Let us consider an abelian (classical) dynamical system $(X, \Sigma, \mu, T)$
where $X$ is a topological Hausdorff space, $\Sigma$ stands for the $\sigma$-algebra of all
Borel subsets of $X$ and $\mu$ is a Borel measure on $X$. $T$ will denote
the automorphism of the measurable space, i.e. $T: X \to X$ is a measurable transformation such 
that both $T$ and $T^{-1}$ are measure preserving.
We recall that in Physics the set of observables $\c O$ is assumed to form a $C^*$-algebra with 
identity. Therefore, taking into account the Gelfand-Naimark theorem
about the structure of abelian $C^*$-algebras, from now on we make the assumption 
that $X$ is a compact space. We use to denote $C_{\Cn}(X)$ ($C_{\Rn}(X) \equiv C(X)$)
the complex-valued (real) continuous functions over $X$.
Let $\phi$ be a positive linear (normalized) functional on $C(X)$. 
The measure-theoretic approach views $\phi$
as ``integration'' relative to an associated positive (probability) measure $\mu$ on $X$ (through 
the Riesz representation theorem).

Turning to states over a general set of observables $\cO$ it is convenient,
as it was mentioned,  to assume that 
$\cO$ generates the $C^*$-algebra $\cM$ with unit. 
The set of all states (linear, positive, normalized functionals) over $\cM$ will 
be denoted by $\cS(\cM) \equiv \cS$. Further, we recall that
any density matrix (positive operator of trace equal to $1$) on $\cal H$ determines 
uniquely a linear
positive, normalized, functional $\omega_{\varrho}(\cdot) \equiv \omega (\cdot)
\equiv Tr\{ \varrho \cdot \}$ on ${\cal B}({\cal H})$ which is also called 
a normal state.
We will assume the Ruelle's separability condition for $\cM$ (cf. \cite{Ru1},
\cite{Ru2}, \cite{BR}): a subset $\cF$ of the set of all states $\cS$ 
of $\cM$ satisfies separability condition if there exists a sequence
$\{ \cM_n \}$ of sub-$C^*$-algebras of $\cM$ such that 
$\cup_{n \ge 1} \cM_n$ is dense in $\cM$, and each $\cM_n$ contains
a closed, two-sided, separable ideal $\cI_n$ such that
\begin{equation}
\cF = \{ \omega; \omega \in \cS, ||\omega|_{\cI_n} || = 1, n\ge 1 \}
\end{equation}

We recall that this condition leads to a situation 
in which the subsets of states have good measurability properties (cf \cite{BR}). 
Furthermore, one can easily verify that this separability
condition is satisfied if we restrict ourselves to the set of 
normal states on $\cM$ or $\cM$ is a separable $C^*$-algebra.
A $C^*$-algebra with family of states satisfying the separability condition leads
to important class of non-commutative (quantum) dynamical systems. We recall,
a non-commutative (quantum) dynamical system is a
triple $(\cM, \alpha, \phi)$ where $\cM$ is a $C^*$-algebra, $\alpha$ is an 
automorphism over $\cM$, and finally $\phi$ is $\alpha$-invariant state on $\cM$, 
i.e., $\phi \circ \alpha = \phi$.

Now, for the convenience of the reader, we introduce some terminology
and give a short resum\'e of results from convexity and Choquet
theory that we shall need in the sequel
(for details see \cite{phel}, \cite{Alf}, \cite{skau}, \cite{Mey}, and \cite{BR}). Let 
$\cM$ stand for a $C^*$-algebra. From now on, for simplicity of our exposition, 
we make the assumption
of separability for $\cM$. We recall that $\cS$ 
(the state space of $\cM$) is a compact convex set in the $^*$-weak topology.
Further, we denote by $M_1(\cS)$ the set of all probability
Radon measures on $\cS$. It is well known that
$M_1(\cS)$ is a compact subset of the vector space of real, regular
Borel measures on $\cS$. Further, let us recall
the concept of barycenter $b(\mu)$ of a measure $\mu \in M_1(\cS)$:
\begin{equation}
b(\mu) = \int d\mu (\varphi) \varphi
\end{equation}
where the integral is understood in the weak sense. The set
$M_{\omega}(\cS)$ is defined as a subset of $M_1(\cS)$
with barycenter $\omega$, i.e.
\begin{equation}
M_{\omega}(\cS) = \{ \mu \in M_1(\cS), b(\mu) = \omega \}
\end{equation}
$M_{\omega}(\cS)$ is a convex closed subset of $M_1(\cS)$, hence
compact in the weak $^*$-topology.
Thus, it follows by the Krein-Milman theorem that there are "many"
 extreme points in $M_{\omega}(\cS)$. We say 
the measure $\mu$ is simplicial
if $\mu$ is an extreme point in $M_{\omega}(\cS)$. The set of all simplicial 
measures in $M_{\omega}(\cS)$ will be denoted by $\cE_{\omega}(\cS)$.

Further, we will need the concept of orthogonal measures.
To define that concept one introduces firstly the notion of orthogonality
of positive linear functionals on $\cM$: given
positive functionals $\phi, \psi$ on $\cM$
we say that $\phi$ and $\psi$ are othogonal, in symbols, $\phi \bot \psi$,
if for all positive linear functionals $\gamma$ on $\cM$, $\gamma \le \phi$
and $\gamma \le \psi$ imply that $\gamma = 0$.

Turning to measures, let $\mu$ be a regular non-negative
Borel measure on $\cS$ and let $\mu_V$ denote the restriction
of $\mu$ to $V$ for a measurable set $V$ in $\cS$, i.e.
$\mu_V(T) = \mu(V \cap T)$ for $T$ measurable in $\cS$.
If for all Borel sets $V$ in $\cS$ we have
\begin{equation}
\int_{\cS} \varphi d\mu_V(\varphi) \quad \bot 
\quad \int_{\cS} \varphi d\mu_{\cS \setminus V}(\varphi)
\end{equation}
we say that $\mu$ is an orthogonal measure on $\cS$. We recall
that the set of all othogonal measures on $\cS$
with barycenter $\omega$, $O_{\omega}(\cS)$, forms a subset
(in general proper) of $\cE_{\omega}(\cS)$, i.e.
$O_{\omega}(\cS) \subset \cE_{\omega}(\cS)$.

In the set of all probability Radon measures on $\cS$, $M_1(\cS)$, one can define
the order relation $\succ$, indroduced by Choquet,  by saying that
$\mu \succ \nu$ if and only if 
$$ \mu(f) \ge \nu(f)$$
for all continuous, real-valued convex functions on $\cS$. Then, one can prove that $\succ$ is
partial ordering. Moreover, for each $\omega \in \cS$ there is a measure $\mu \in M_{\omega}(\cS)$
which is maximal for the order $\succ$. Furthermore, maximal measures are pseudosupported
(supported if $\cS$ is metrizable or $\omega$ is in a face satisfying separability 
condition) on extremal points $\cE xt (\cS)$ of $\cS$.

\section{Kolmogorov-Sinai entropy}
Let $(X,T, \mu)$ be a classical dynamical system and $\{X_i \}_{i=1}^n$ a partition of $X$, i.e. each $X_i$ is a 
measurable non-empty subset of $X$ such that $X_i \cap X_j = \emptyset$ for $i \ne j$, and
$\bigcup_i X_i = X$. Let $\phi$ be a linear positive normalized functional over $C(X)$ 
associated with the probability measure $\mu$ on $X$ via the Riesz 
representation theorem. Clearly, $T$-invariance of $\mu$ implies the analogous property for
$\phi$, i.e., $\phi \circ U_T = \phi$, where $U_T$ stands for the Koopman's operator.
Further, let us consider $\phi_i^0 \equiv \phi_{\mu_i}$ where $\mu_i \equiv \mu|_{X_i}$, i.e.
$\phi_i^0$ is a linear, positive functional associated with the measure $\mu_i$.
We have
\begin{equation}
\label{rozklad1}
\phi = \sum_{i=1}^n \phi_i^0 = \sum_{i=1}^n \phi_i^0({\bf 1}_X) 
{ \phi_i^0 \over \phi_i^0({\bf 1 }_X)}
\equiv \sum_{i=1}^n \phi_i^0({\bf  1}_X) \phi_i
\end{equation} 
where ${\bf 1}_X$ stands for identity function on $X$ so it is 
the unit $\idty$ of the (abelian) algebra
$C(X)$. Next, let us observe that the condition $\psi \le \phi_i$ and $\psi \le \phi_j$
with $i \ne j$ 
for a positive functional
$\psi$ implies that $\psi = 0$. Thus, we got in (\ref{rozklad1}) 
an orthogonal finite decomposition
of the state $\phi$ (we repeat, the state $\phi$ is 
associated with the probability measure $\mu$).
Consequently, we got a hint that in algebraic reformulation of definition of Kolmogorov-Sinai
entropy it is convenient to replace the concept of (finite) partition by the concept of
(finite) orthogonal decomposition of the corresponding state. 
More precisely, let us denote by $\cS_c$ the state space of $C(X)$ (its extremal
points $\cE xt(\cS_c)$ can be 
identified with $X$). We define the measure $\nu_{\phi} \in M_{\phi}(\cS_c)$ as
\begin{equation}
\label{rozklad2}
\nu_{\phi} = \sum_{i=1}^n \phi_i^0{(\idty)} \delta_{\phi_i}
\end{equation}
where $\delta_{\phi_i}$ stands for the Dirac (point) measure. Thus we are replacing
the partition $\{ X_i \}$ by the orthogonal measure $\nu_{\phi}$.

To go further, let us recall some basic facts from the theory of representation
of operator algebras (for all necessary details see \cite{BR}, \cite{Kad}, or \cite{tak}).
The GNS triple associated with the pair $(C(X), \phi)$ can be identified with
\begin{equation}
( \cH_{\mu} = \cL^2(X, \mu), (\pi_{\mu}(f) \xi)(x) = f(x) \xi(x), \Omega_{\mu} = {\bf 1}_X)
\end{equation}
where $\xi \in \cH_{\mu}$, $x \in X$, and finally ${\bf 1}_X$ stands for the identity 
function on $X$. Furthermore, the von Neumann algebra generated by $\pi_{\mu}(C(X))$
is maximal abelian one and it
can be identified with the algebra $\cL^{\infty}(X, \mu)$ of all 
essentially bounded function on $X$. On the other hand, there is one-to-one
correspondence between an othogonal decomposition of a state $\phi$ and abelian
subalgebra in $\pi_{\mu}(C(X))^{\prime}$.
Let us describe that abelian algebra in some details. 
Let $\chi_{_{X_i}} \equiv \chi_i$ stands for the 
characteristic function associated with the subset $X_i$. The (abelian) algebra
generated by $\{ \chi_i \}_{i =1}^n$ will be denoted by $\cA_0$.
Denote by $P$ the projector of $\cL^2(X, \mu)$ onto $\overline{ \cL^{\infty}(X, \mu) {\bf 1}_X}$.
$\cA_0$ maps $\overline{ \cL^{\infty}(X, \mu) {\bf 1}_X}$ into 
$\overline{ \cL^{\infty}(X, \mu) {\bf 1}_X}$ and $P$ is its unit.

By the above and the characterization of orthogonal measures in terms of abelian algebras
(see \cite{skau}, {\cite{BR}) we have the one-to-one correspondence between
the partition $\{ X_i \}$, the orthogonal measure $\nu_{\phi}^{(\cA_0)}$ with fixed barycentre $\phi$
and the abelian von Neumann algebra $\cA_0$ in the corresponding commutant.

Now, let us take into account the dynamic map $T$. We define

$$ \cA_1 \equiv algebra\{ \chi_{_{X_1}}, ...,\chi_{_{X_n}}, \chi_{_{T(X_1)}},
...,\chi_{_{T(X_n)}} \}$$
$$.$$
$$.$$
$$.$$
$$\cA_k \equiv algebra \{ \chi_{_{X_1}}, ..., \chi_{_{X_n}}, \chi_{_{T(X_1)}},
...,\chi_{_{T(X_n)}},..., \chi_{_{T^k(X_1)}},..., \chi_{_{T^k(X_n)}}\}$$
where $algebra \{ a,b,c,.. \}$ stands for the $W^*$ algebra generated by $a, b, c, ..$.
Clearly, $\cA_l \subseteq \cA_k$ for $l \le k$. This is very important, and we will need it later,
as there is the following equivalence (cf. \cite{skau}, or \cite{BR}):
\begin{equation}
\label{rozklad3}
\nu_{\phi}^{(\cA_k)} \succ \nu_{\phi}^{(\cA_l)} \leftrightarrow  \cA_k \supseteq \cA_l
\end{equation}
where $\mu_{\phi}^{(\cA_k)}$ stands for the orthogonal measure on $\cS_c$ uniquely determined by $\cA_k$,
while the relation $\succ$ is the Choquet's relation (cf Section 2).

Now, let us turn to construction of KS entropy. We have fixed a dynamical system
$(X,T, \mu)$, so we fixed an abelian $C^*$-algebra $C(X)$ and a state $\phi$ over it.
Then, we take a finite, orthogonal decomposition of the state $\phi$ determined by 
the abelian algebra $\cA_0$ generated by mutually othogonal projectors $\{ \chi_i \}$. Subsequently, 
we form a sequence of abelian algebras
$\cA_k$ generated by projectors $\{\{ \chi_{_{T^p(X_i)}} \}_{p=0}^k \}_{i=1}^n$. 
Taking the evaluation of the state $\phi$ on 
$\chi_{_{T^p(X_i)\cap T^r(X_j) }}$,  $\phi(\chi_{_{T^p(X_i) \cap T^r(X_j)}}) \equiv y_{i,j,p,r}
\equiv y_{\bf a}$, 
we associate with each algebra $\cA_k$
the number
\begin{equation}
H_{\mu, T}(\cA_k) = \sum_{\bf a}  \eta(y_{\bf a})
\end{equation}
where $\eta$ stands for the function $x \mapsto \eta(x) = - xlnx$.

Then, the KS entropy is defined as

\begin{equation}
h_{\mu}(T) = sup_{\nu_{\phi}^{(\cdot)}} lim_k H_{\mu, T}(\cA_k)
\end{equation}
where the $sup$ is taken over all finite orthogonal probability measures 
$\nu_{\phi}^{(\cdot)}$ with fixed 
barycenter $\phi$. We recall that each measure $\nu_{\phi}^{(\cdot)}$ uniquely
corresponds to a finite partition, i.e. $\nu_{\phi}^{(\cA_0)}$ corresponds to a finite partition
associated with abelian von Neumann algebra $\cA_0$.
Clearly, the above definition is just a reformulation of 
the original one, that given by Kolmogorov.

We want to close this section with 

\begin{rems}
\begin{enumerate}

\item For the classical case, considered in this Section, the state space forms a simplex. 
This implies (cf. \cite{Alf}) that the set of $M^0_{\phi}(\cS_c)$ of finite probability measures
with fixed barycenter $\phi$ is directed in ordering of Choquet.
Consequently, the orthogonal measure $\mu_k$ corresponding to the algebra $\cA_k$
is the smallest one from the set of probability measures with fixed barycenter $\phi$ and 
majorizing the measure $\mu_{k-1}$ determined by $\cA_{k-1}$ and the measure determined
by the algebra generated by $\{ \chi_{_{T^k(X_1)}}, ..., \chi_{_{T^k(X_n)}} \}$.

\item If $\mu$ is a Dirac measure, 
then the prescription for dynamical entropy
is trivial. We wish to have the same property for the quantum case.

\item Let the increasing sequence of algebras $\{ \cA_k \}$ generate the maximal abelian algebra.
Then, the calculation of KS entropy simplifies significantly. That case
corresponds to Kolmogorov-Sinai theorem about the generator.

\end{enumerate}
\end{rems}

\section{Quantization of dynamical entropy}
Let us consider a non-commutative dynamical system $(\cM, \alpha, \phi)$
where $\cM$ is a $C^*$-algebra, $\alpha$ is an automorphism over $\cM$ and
$\phi$ is a $\alpha$-invariant state on $\cM$, i.e. $\phi \circ \alpha = \phi$.
As, in Section 3, we form GNS triple $(\cH_{\phi}, \pi_{\phi}, \Omega_{\phi})$
associated with $(\cM, \phi)$. As $\alpha : \cM \to \cM$ is $\phi$-invariant
automorphism then, there is  the unitary operator $U : \cH_{\phi} \to \cH_{\phi}$
such that $\pi_{\phi}(\alpha(A))\Omega_{\phi} = U \pi_{\phi}(A) \Omega_{\phi}$.
Denote by $\alpha^0$ the corresponding automorphism over $\cB(\cH_{\phi})$,
i.e., $\alpha^0(A) = UAU^*$, for $A \in \cB(\cH)$.
We observe that $\alpha^0(\pi_{\phi}(\cM))^{\prime} \subset \pi_{\phi}(\cM)^{\prime}$.
Consequently, $\alpha^0$ maps any abelian von Neumann subalgebra $\cA \subseteq 
\pi_{\phi}(\cM)^{\prime}$ into the abelian subalgebra of $\pi_{\phi}(\cM)^{\prime}$.

In order to obtain the announced prescription for quantum Kolmogorov-Sinai
entropy we first examine the simplest situation in which $\pi_{\phi}(\cM)^{\prime}$
is abelian. Then, we will pass to the general case.

\subsection{Multiplicity-free representation $\pi_{\phi}$}
We recall (see \cite{skau}) that a representation $\pi_{\phi}$ of $\cM$ is said to be
multiplicity-free if $\pi_{\phi}(\cM)^{\prime}$ is abelian. 
Having that property we can repeat the definition of KS-entropy
given in the previous Section. That is, we take a finite othogonal measure
$\nu_{\phi}$ in $M_{\phi}(\cS(\cM))$. By the general correspondence 
(described in previous Sections),
with that measure is (uniquely) associated abelian (finite) subalgebra $\cA_0
\in \pi_{\phi}(\cM)^{\prime}$. Take $\alpha^0(\cA_0)$. This is abelian, finite
subalgebra in $\pi_{\phi}(\cM)^{\prime}$. Let $\cA_1$ be the von Neumann algebra
generated by $\cA_0$ and $\alpha^0(\cA_0)$. As $\cA_0$ and $\alpha^0(\cA_0)$
 are finite subalgebras in the abelian algebra then $\cA_1$ is finite abelian von Neumann
subalgebra in $\pi_{\phi}(\cM)^{\prime}$. Denote by $\mu_1 \in O_{\phi}(\cS)$ 
the corresponding orthogonal measure. Clearly, a repetition of that procedure 
leads to the succeeding measure $\mu_2$, 
etc. Consequently, with each measure $\mu_k$ (so with the algebra $\cA_k$) 
we can associate the number $H_{\mu, \alpha}(\cA_k)$ by the same rule as before.  
As the rest is evident this finishes the quantization of KS-entropy for that case.

We want to close this subsection with the following observation. Let $\phi$ be a pure 
state on $\cM$. Then, $\pi_{\phi}(\cdot)$ is the irreducible representation.
Hence, $\pi_{\phi}(\cM)^{\prime} = \{ \lambda \idty \}$. Therefore, there is only
one abelian von Neumann algebra in $\pi_{\phi}(\cM)^{\prime}$. Thus, the procedure
determining the dynamical entropy becomes trivial and we have the answer
to the question posed in Remarks 1.2.

\subsection{General case}
As in general case, $\pi_{\phi}(\cM)^{\prime}$ does not need to be abelian (e.g. if $\phi$ would 
stand for the KMS (quantum Gibbs state)) there is a difficulty in carrying
out directly the above construction. Namely, even in the first step, the von Neumann algebra 
generated by $\cA_0$ and $\alpha^0(\cA_0)$ does not need to be abelian.
To overcome that problem we will proceed as follows.
Let $\mu^0$ be a measure in $M_{\phi}(\cS)$. Further, let $\cP_0 = \{ Y_i \}_{i=1}^n$
be a partition of $\cS$. Let $\chi_i$ denote the characteristic function of $Y_i$
and define $\lambda_i$ and $\mu^0_i$ by $\lambda_i = \mu^0(Y_i)$ and
$\lambda_id\mu^0_i = \chi_id\mu^0$. Thus
\begin{equation}
\mu^0 = \sum_{i=1}^n \lambda_i \mu^0_i
\end{equation}
As, for probability Radon measure $\sigma \in M_1(\cS)$ 
there exists the unique barycenter, one has
existence of states $\phi_i$ such that $\mu_i^0 \in M_{\phi_i}(\cS)$
and
$$\phi = \sum_{i=1}^n \lambda_i \phi_i.$$
Put
\begin{equation}
H_{\phi,\mu^0, \alpha}(\cP_0) = - \sum_i \lambda_i ln\lambda_i
\end{equation}
As, $\alpha$ is an automorphism of $\cM$, there is the weak $^*$-continuous
affine isomorphism $T_{\alpha}: \cS \to \cS$ such that
\begin{equation}
(T_{\alpha} \phi)(B) = \phi \circ \alpha(B)
\end{equation}
where $B \in \cM$ (cf \cite{Kad1}).
Define $\cP_1 = \{ Y_i \cap T_{\alpha}(Y_j) \}_{i,j}$ and repeat the above procedure,
 now leading to $H_{\phi,\mu^0, \alpha}(\cP_1) $. Thus, we arrive to the sequence 
of partitions $\cP_k$, the sequence of decompositions of state $\phi$
\begin{equation}
\phi = \sum_i^{(k)} \lambda_i^{(k)} \phi_i^{(k)},
\end{equation}
and to the sequence of numbers $H_{\phi,\mu^0, \alpha}(\cP_k) $.
We put
\begin{equation}
H_{\phi,\mu^0, \alpha}(\cP) = \lim_k H_{\phi,\mu^0, \alpha}(\cP_k) 
\end{equation}
and we define
\begin{equation}
h_{\phi}(\alpha) = \sup_{\mu^0 \in O_{\phi}(\cS)} \sup_{\cP} H_{\phi,\mu^0, \alpha}(\cP) 
\end{equation}
where $O_{\phi}(\cS)$ stands for the set of all orthogonal measures with barycenter
$\phi$. {\it $h_{\phi}(\alpha)$ is the quantized K-S entropy 
for non-commutative dynamical system $(\cM, \alpha, \phi)$.}

\section{Remarks}
Let $(\cM_i, \alpha_i, \phi_i)$, $i=1,2$ be two isomorphic dynamical systems, i.e., 
there is an isomorphism $\beta$ such that $\beta: \cM_1 \to \cM_2$ and 
$\beta \circ \alpha_1 = \alpha_2 \circ \beta$. Our assumption implies that two dynamical
systems
\begin{equation}
(\cS(\cM_i), T_{\alpha_i}, \mu^{0,i}),
\end{equation}
for properly chosen measures $\mu^{0,1}$ and $\mu^{0,2}$ are also isomorphic.
Here, properly chosen measures means that picking up the measure $\mu^{0,1}$ for the first system, 
the measure $\mu^{0,2}$ is determined by the equality $\mu^{0,2} = \mu^{0,1} \circ T_{\beta}$
where $T_{\beta}$ is defined by $(T_{\beta}\varphi)(A) = \varphi(\beta(A))$ for $A \in \cM_1$,
$\varphi \in \cS(\cM_2)$. Therefore, 
$\sup_{\cP} \lim_k H_{\phi_i,\mu^{0,i}, \alpha_i}(\cP_k^i)$
are equal to each other by the classical Kolmogorov theorem.
As the rest is clear, we arrived to : {\it $h_{\phi}(\alpha)$ is a dynamical-system invariant.}

Next, let us compare definitions of dynamical entropy for commutative and non-commutative
dynamical system. The main difference between both cases is that we have additional
$sup$ over measures in $O_{\phi}(\cS)$ for the latter case. The reason for that is clear.
Namely, for an abelian case, $\cS_c$ forms simplex and as the ``best''
decomposition is determined by the (unique) measure (supported by $\cE xt(\cS_c) \equiv X$),
we have a kind of uniqueness in the recipe for dynamical entropy. 
That feature does not hold for non-commutative case. Thus, we were
forced to consider all measures in $O^0_{\phi}(\cS)$.

Clearly, in (16) one can replace $\sup_{\mu \in O_{\phi}}$ by $\sup_{\mu \in M_{\phi}}$.
The result would be another dynamical-system invariant 
$h^{\prime}_{\phi}(\alpha) \ge h_{\phi}(\alpha)$. Our choice of $\sup_{\mu \in O_{\phi}}$
was motivated by the role of orthogonal measures in the definition of K-S
entropy for abelian case as well as for multiplicity-free representations.

To summarize: we present the concise definition of non-commutative dynamical entropy
without refering to additional, supplementary structures and constructions
(auxiliary abelian systems, smooth elements, etc.). Moreover, the presented definition
is given within the same scheme which was used to analize quantum entanglement and to 
define quantum correlations.

\end{document}